\documentclass[twocolumn,prb,showpacs,floatfix]{revtex4}

\usepackage{graphicx}
\usepackage{bm}
\usepackage{epstopdf}
\usepackage{amsmath,amsfonts,paralist}
\usepackage{mathrsfs}
\usepackage{eufrak}

\let\a=\alpha \let\b=\beta  \let\g=\gamma  \let\d=\delta \let\e=\varepsilon
     \let\l=\lambda
\let\m=\mu    \let\n=\nu         \let\p=\pi   
\let\s=\sigma \let\t=\tau    
   
\let\G=\Gamma \let\D=\Delta

 \def\HH{{\cal H}}

\def\GG{{\cal G}} \def\SS{{\cal S}}
\def\UU{{\cal U}}

\def\ul{\underline}

\def\ol{\overline}
\def\ox{\overline{x}}

\def\to{\rightarrow}

\newcommand{\beq}{\begin{equation}}
\newcommand{\eeq}{\end{equation}}

\newcommand{\Tr}{\text{Tr}}

\begin{document}

\title{Dynamical critical exponents for the mean-field Potts glass}


\author{F.  Caltagirone$^{1,2}$,
G. Parisi$^{1,2,3}$ and T. Rizzo$^{1,2}$} \affiliation{$^1$ Dip. Fisica,
Universit\`a "Sapienza", Piazzale A. Moro 2, I-00185, Rome, Italy \\
$^2$ IPCF-CNR, UOS Rome, Universit\`a "Sapienza", PIazzale A. Moro 2,
I-00185, Rome, Italy \\ $^3$ INFN, Piazzale A. Moro 2, 00185, Rome,
Italy}

\begin{abstract}
In this paper we study the critical behaviour of the fully-connected $p$-colours Potts spin-glass at the dynamical transition. 
In the framework of Mode Coupling Theory (MCT), the time autocorrelation function displays a two step relaxation, with two exponents 
governing the approach to the plateau and the exit from it. 
Exploiting a relation between statics and equilibrium dynamics which has been recently introduced, we are 
able to compute the critical slowing down exponents at the dynamical transition with arbitrary precision and for any value of the number of colours $p$. When available, we compare our 
exact results with numerical simulations. In addition, we present a detailed study of the dynamical transition in the large $p$ limit, showing 
that the system is not equivalent to a random energy model.

\end{abstract}
\date{\today} \maketitle

\section{Introduction }
\label{sec:intro}
Mean-field spin-glass models can be divided into two main classes, the ones which undergo a continuous transition and the ones 
which, instead, display a jump in the order parameter.
In systems belonging to the former class, at a certain temperature $T_s$ a second order phase transition takes place, 
with a continuous growth of the Edwards-Anderson order parameter $q_{EA}=\frac{1}{N}\sum_i \ol{\langle S_i\rangle^2}$ and 
zero magnetization (in absence of magnetic field): the spins are essentially frozen in a random direction so that the global 
mean magnetization vanishes while the mean squared magnetization is finite. In the low temperature phase the replica symmetry
is broken with a continuous pattern (Full RSB) or with a step-like pattern (1-RSB) according to the Parisi scheme \cite{Parisi80} and the order parameter is, in fact, a 
non-trivial function $q(x)$.\\
One can also study the Langevin dynamics of these systems, showing that exactly at the thermodynamic transition temperature $T_s$ there is 
ergodicity breaking, therefore we can say that, in correspondence of the static transition, a dynamical transition takes place too. \\ 
There exists another class of mean-field spin-glass models (like the $p$-spin or the $p$-colours Potts model \cite{Crisanti92, Gross85}) which display two different transitions: at a temperature $T_s$ there is a thermodynamic phase transition which is of the second order in terms of potentials but 
can be discontinuous in the EA order parameter. The low temperature phase is (at least in the vicinity of the critical temperature) 1-step replica symmetry broken. 
At a temperature $T_d > T_s$ a dynamical phase transition occurs, where the system's relaxation time becomes infinite 
and the ergodicity is broken \cite{Crisanti92, Crisanti93}. This is due to the fact that at the dynamical transition the equilibrium state splits into 
a large (exponential in the system size) 
number of excited states, represented by free energy local minima. Since in mean-field the barriers between these states become infinitely high 
in the thermodynamic limit, the equilibrium dynamics remains stuck forever in one of them and
the overlap cannot relax to zero.\\
This second class of mean-field systems has been shown to share some relevant properties of 
structural glasses \cite{Kirkpatrick87a, Kirkpatrick87b, Kirkpatrick87c, Kirkpatrick87d}, more specifically, 
the dynamical equations are exactly equivalent to those predicted by 
the Mode Coupling Theory (MCT) above the mode coupling temperature $T_{mc}$ where ergodicity breaking occurs. 
The analogy between structural glass models (with self-induced frustration) and proper mean field spin-glasses (with quenched disorder) has been widely studied and has provided rather accurate predictions \cite{Monasson95, Mezard96, Mezard99a, ParZamp10} .
In systems with continuous transition, above $T_s$ the spin-spin time correlation function $C(t)=\langle \sigma_i (0) \sigma_i (t)\rangle$ decays exponentially at large times, which means that the system is ergodic. Lowering the temperature the relaxation time grows until it diverges exactly at $T_s$ (the static transition temperature) so that the ergodicity is broken and the relaxation (at large times) follows a power law $C(t)\sim t^{-\nu}$ with some exponent $\nu$.\\
The systems belonging to the discontinuous class introduced above, behave quite differently: above $T_d$ the time correlation function displays at first a fast decay to a plateau and then a slow decay to zero (in absence of a magnetic field) \cite{Crisanti93};  the length of the plateau grows lowering the temperature until it diverges at $T_d$. \\ 
According to MCT the approach to the plateau and the decay from it are both characterized by a power-law behaviour, respectively
\beq
\label{eqa}
C(t)\simeq q_d + ct^{-a} 
\eeq
\beq
\label{eqb}
C(t) \simeq q_d - c' t^{b}
\eeq
where $q_d$ is the height of the plateau and the two exponents satisfy the exact MCT relation
\beq
\frac{\Gamma^2(1-a)}{\Gamma(1-2a)}=\frac{\Gamma^2(1+b)}{\Gamma(1+2b)}=\l
\eeq
and $\l$ is usually treated as a tunable parameter (see for example \cite{Gotze}). \\
The exponents $a$ and $b$ have been computed exactly only for the spherical $p$-spin model \cite{Crisanti93} because the 
dynamical equations are particularly simple and correspond to the so called schematic MCT models.\\
In most of the cases it is instead very difficult or impossible to compute the exponents in a purely dynamical framework, both 
analytically or through Monte Carlo simulations. \\
Numerical simulations are often difficult to interpret and give quite poor indication of the value of the exponents due 
to strong finite size effects; if the system is not infinite, barriers between metastable states cannot be infinitely high, the dynamics does 
not remain stuck in a single state and all the observables eventually relax to their equilibrium value since, through activated processes, the configuration
is able to explore the whole phase space. 
The extent of this effect depends on the specific model we consider and, in particular, on how fast the barriers between metastable states grow
with the size of the system. \\
Recently, a connection between the mode coupling exponents and some purely thermodynamic quantities has been introduced \cite{Calta11};
this connection suggests a quite simple recipe to compute the dynamical exponents exactly starting from the static mean-field theory 
which is much easier to work out for all the reasonable models one can think of. \\
The aim of this paper is to apply this technique to the mean-field Potts spin-glass and compute the MCT exponents for 
any value of the number of colours $p$.
The Potts glass is particularly interesting because, as will be pointed out in the following sections, the 
parameter $p$ allows to switch from a continuous transition ($p \leq 4$) to a discontinuous transition ($p>4$), moreover in the latter case it works 
as a tuning parameter for the magnitude and separation of the static and dynamical transitions.
In the following we will not make use of the symplectic representation which is widely exploited in literature \cite{Wallace75, Elderfield83a, Elderfield83b}.
The outline of the paper is the following: in section \ref{sec:how} we give a sketch of the technique used to compute the MCT exponents in a generic model, in section \ref{sec:summary} we summarize some of the necessary known results about the Potts model, in section \ref{sec:pottsexp} we compute the dynamical exponents for the Potts model for arbitrary value of the parameter $p$, in section \ref{sec:simul} we compare our theoretical exact results with numerical simulations and, finally, in section 
\ref{sec:conclusion} we give our conclusions and final remarks.

\section{How to compute the exponent}
\label{sec:how}

Given a fully-connected model it is possible to compute the Gibbs free energy as a function of the order parameter which, 
in the case of a spin-glass transition is the well known overlap matrix $Q$. The thermodynamic value of the order parameter can be determined minimizing the Gibbs free energy functional. It can then be expanded around the replica symmetric saddle point solution, giving raise to eight different kinds of third order terms. For our purposes, only two of them will be relevant, namely:
\beq
w_1 \Tr (\delta Q^3)= w_1 \sum_{a,b,c} \delta Q_{ab} \delta Q_{bc} \delta Q_{ca}
\eeq
and
\beq
w_2 \sum_{a,b} \delta Q_{ab}^3
\eeq
In the case of continuous transitions it has been found \cite{Calta11} that there exists a quite simple relation between the exponent $\nu$ and the two coefficients $w_1$ and $w_2$: 
\beq
\frac{\Gamma^2(1-\nu)}{\Gamma(1-2\nu)}=\frac{w_2(T_s)}{w_1(T_s)}
\label{rel}
\eeq
In the case of discontinuous transitions it can be shown \cite{Calta11} that a relation analogous to (\ref{rel}) holds at the dynamical transition which, again, gives the connection between the dynamical exponents $a$ and $b$ and the static coefficients, namely
\beq
\frac{\Gamma^2(1-a)}{\Gamma(1-2a)}= \frac{\Gamma^2(1+b)}{\Gamma(1+2b)} = \frac{w_2(T_d)}{w_1(T_d)}
\eeq 
where, differently from the former case, the expansion of the Gibbs free energy has to be performed around the dynamical overlap (the height of the infinite plateau at the dynamical transition). \\
In order to compute the two coefficients $w_1$ and $w_2$ one must determine the expression 
of the Gibbs free energy as a function of the overlap and then expand it to third order around the RS
thermodynamic value $q$.
The reason why the expansion has to be performed around a replica symmetric solution will be clarified in section \ref{sec:pottsexp}.
In fully connected models, introducing a replicated external field $\e$, the free energy reads
\beq
\begin{split}
f(\e)=-\frac{1}{\b n N} \ln \int dQ \exp N\left( \SS[Q] + \Tr \, \e Q\right) 
\end{split}
\eeq
which, for $N \to \infty$, can be evaluated at the saddle point
\beq
\begin{split}
f(\e)=-\frac{1}{\b n } \mathrm{extr}_Q  \left( \SS[Q] + \Tr \, \e Q\right) 
\end{split}
\eeq
We can immediately notice that the equation above exactly defines $f(\e)$ as the {\it Anti Legendre Transform} ($\ol{\mathscr{L}}$) of the effective action
\beq
\begin{split}
f(\e)=\ol{\mathscr{L}} ( \SS[Q])
\end{split}
\eeq
and, again, by definition the Gibbs free energy $\G(Q)$ is the {\it Legendre Transform} ($\mathscr{L}$) of $f(\e)$, yielding
\beq
\begin{split}
\G(Q)\equiv\mathscr{L} ( f(\e))= \mathscr{L}  \left( \ol{\mathscr{L}} (\SS[Q])\right) =\SS[Q]
\end{split}
\eeq
This implies that the functional form of the Gibbs free energy is exactly the same of the effective action. In fully connected models, 
we can then directly expand the latter.\\
The general form of the third order term in the effective action reads
\beq
\SS^{(3)} = \sum_{(ab)(cd)(ef)} W_{ab,cd,ef} \, \d Q_{ab} \d Q_{cd} \d Q_{ef}
\label{third}
\eeq
Since $a\neq b$, $c\neq d$ and $e \neq f$ and the coefficients $W$ are 
computed in RS {\it ansatz}, we can have eight different vertices: 
\beq
\begin{split}
&W_{\a\b,\b\g,\g\a}=W_1 \,\, , \,\,
W_{\a\b,\a\b,\a\b}=W_2 \\
&W_{\a\b,\a\b,\a\g}=W_3 \,\, , \,\,
W_{\a\b,\a\b,\g\d}=W_4 \\
&W_{\a\b,\b\g,\g\d}=W_5 \,\, , \,\,
W_{\a\b,\a\g,\a\d}=W_6 \\
&W_{\a\g,\b\g,\d\m}=W_7\,\, , \,\,
W_{\a\b,\g\d,\m\n}=W_8 .
\end{split}
\eeq
Following Ref. \cite{temesvari}, Eq. (\ref{third})
can be rephrased in the following way
\beq
\begin{split}
\SS^{(3)} &= w_1\sum_{\a\b\g}\d Q_{\a\b}
\d Q_{\b\g}
\d Q_{\g\a}+w_2\sum_{\a\b}\d Q_{\a\b}
\d Q_{\a\b}
\d Q_{\a\b}\\
&+w_3\sum_{\a\b\g}\d Q_{\a\b}
\d Q_{\a\b}
\d Q_{\a\g}
+w_4\sum_{\a\b\g\d}\d Q_{\a\b}
\d Q_{\a\b}
\d Q_{\g\d} \\
&+w_5\sum_{\a\b\g\d}\d Q_{\a\b}
\d Q_{\a\g}
\d Q_{\b\d}
+w_6\sum_{\a\b\g\d}\d Q_{\a\b}
\d Q_{\a\g}
\d Q_{\a\d} \\
&+w_7\! \! \! \!  \sum_{\a\b\g\d\m}\d Q_{\a\g}
\d Q_{\b\g}
\d Q_{\d\m}
+w_8 \! \! \! \! \sum_{\a\b\g\d\m\n}\d Q_{\a\b}
\d Q_{\g\d}
\d Q_{\m\n}
\end{split}
\label{pimentel}
\eeq
with
\beq
\begin{split}
w_1&=W_1-3W_5+3W_7-W_8\\
w_2&=\frac{1}{2}W_2-3W_3+\frac{3}{2}W_4+3W_5
+2W_6-6W_7+2W_8\\
w_3&=3W_3-3W_4-6W_5-3W_6+15W_7-6W_8\\
w_4&=\frac{3}{4}W_4-\frac{3}{2}W_7
+\frac{3}{4}W_8\\
w_5&=3W_5-6W_7+3W_8\\
w_6&=W_6-3W_7+2W_8\\
w_7&=\frac{3}{2}W_7-\frac{3}{2}W_8\\
w_8&=\frac{1}{8}W_8
\label{ws}
\end{split}
\eeq
It is therefore sufficient to compute the eight W coefficients and 
use Eq. (\ref{ws}) to get $w_1$ and $w_2$.\\

\section{The Potts Model: Summary of known results}
\label{sec:summary}

We consider the $p$-colours disordered Potts Hamiltonian
\beq
\HH=-\sum_{<i,j>} \, J_{ij} \, \eta(\s _i , \s _j)
\eeq
with
\beq
\eta(a, b)= p \, \delta_{a,b} - 1
\eeq
where $p$ is the number of colours and $\s=0,1, \cdots, p-1$. \\
The sum is extended over all the possible couples taken from $N$ spins and the couplings $J_{ij}$ are 
independent gaussian random variables with mean $J_0/N$ and variance $J^2/N$, where the normalization is needed 
in order to obtain a finite thermodynamic limit.
As usual, we are interested in computing the mean-field free-energy exploiting the well known replica trick in order to average over 
the disorder
\beq
\ol{\ln Z}= \lim_{n \rightarrow 0} \frac{1}{n} \ln \ol{Z^n}
\eeq
Carrying on the computation we obtain the replicated partition function in a functional integral form
\beq
\ol{Z^n}=\int D\ul{Q}\, D\ul{m} \, \exp (-N\SS[m,Q])
\eeq
\noindent where the ``effective action'' $\SS[m,Q]$ is a function of two order parameters:
the magnetization $m_{r}^{\a}$ and the overlap $Q_{rs}^{\a \b}$, with greek replica indices $\a,\b = 1, \cdots , n$ and latin color indices 
$r,s=1, \cdots, p$.
\beq
\begin{split}
&\SS[m,Q]=\frac{\b^2 J^2}{4}(1-p) + \frac{\b^2 J^2}{2p^2}\sum_{\a < \b}\sum_{r,s} (Q^{\a\b}_{rs})^2 \\
& \frac{\b}{2p} \left[ J_0 + \b J^2 \frac{p-2}{2} \right] \sum_{\a}\sum_{r} (m^{\a}_{r})^2  - \ln \Tr_{ \{ \s \} } e^ {\HH[m,Q, \{ \s \}  ] }
\end{split}
\eeq

\beq
\begin{split}
\HH[m,Q, \{ \s \}  ] & = \frac{\b^2 J^2}{p^2}\sum_{\a < \b}\sum_{r,s} Q^{\a\b}_{rs} \, \eta(\s^{\a},r) \eta(\s^{\b},s) \\
& + \frac{\b}{p} \left[ J_0 + \b J^2 \frac{p-2}{2} \right] \sum_{\a}\sum_{r} m^{\a}_{r} \, \eta(\s^{\a},r)
\end{split}
\eeq
In order to determine the order parameters we can use the two saddle point equations, which read
\beq
Q^{ab}_{rs}=\langle\langle \eta(\s^{\a},r) \eta(\s^{\b},s) \rangle\rangle
\eeq

\beq
m^{\a}_r=\langle\langle \eta(\s^{\a},r) \rangle\rangle
\eeq
where $\langle\langle \cdots \rangle\rangle$ is the average taken with respect to the measure 
\beq
\m( \{ \s \} ) = \frac{e^{ \HH[m,Q,\{ \s \}] } }{ \Tr_{ \{ \t \} } e^{ \HH[m,Q,\{ \t \}] } }
\eeq
The order parameters are clearly redundant, in fact they satisfy the following constraints:
\beq \begin{split}
\sum_{r}Q_{r s}^{\a \b} &= 0 \,\,\,\,\, \forall s\\
\sum_{r} m_{r}^{\a} &= 0
\end{split} \eeq
In the particular case $p=2$ one recovers the SK model solution \cite{Parisi80}. \\
For $p>2$ ferromagnetic ordering is always preferred below some temperature $T_{F}$ \cite{Gross85}. 
An upper bound $T_E$ for the temperature $T_F$ below which ferromagnetic ordering appears is \cite{parisiPotts} (from now on we consider $J=1$):
\beq
T_E=\frac{p-2}{2(1 - J_0)}
\eeq
For $p>4$, in order to prevent that ferromagnetic ordering occurs at a higher temperature than the spin-glass one, 
the couplings should be antiferromagnetic in average, with $J_0$ less than some (negative) threshold value. A lower bound 
for the critical mean-value is $J_F=(4-p)/2$ \cite{parisiPotts}.
Under this condition the magnetization is zero and it is straightforward to show that, as a consequence, the overlap has the symmetry 
$Q_{rs}^{\a \b} = Q^{\a \b} \eta(r,s)$. Then it is possible to write the Gibbs free energy as a function of a unique overlap 
matrix in the following way \cite{parisiPotts}:
\beq
\begin{split}
\G[Q] & = \frac{1}{2}(p-1)\beta^2 \sum_{\a < \b} Q_{\a \b}^2 -\\
& - \log \Tr \exp \left[ \beta^2 \sum_{\a < \b} Q_{\a\b} \eta (\sigma^{\a} , \sigma^{\b})\right]
\end{split}
\label{pottsaction}
\eeq
Differentiating with respect to $Q_{\a \b}$ one obtains the saddle point equation
\beq
\begin{split}
Q_{\a \b} & = \frac{1}{p-1} \frac{\Tr \, \eta(\s^{\a}, \s^{\b}) \exp \left(\UU\left[Q,\s\right]\right)}{\Tr \exp \left( \UU\left[Q,\s\right] \right)} = \\
& = \frac{1}{p-1}  \langle \langle  \eta(\s^{\a} , \s^{\b}) \rangle \rangle
\end{split}
\eeq
with
\beq
\UU\left[Q,\s\right]= \b^2  \sum_{\a < \b} Q_{\a \b} \eta (\s^{\a} , \s^{\b})
\eeq
It has been shown \cite{Gross85} that for $2.8<p<4$ the system undergoes a continuous transition at a temperature $T_{s}=1$ with 1-step RSB. The breaking point is 
$\ol{m}=(p-2)/2$. \\
For $p>4$ the transition occours at a temperature $T_{s}>1$, it is discontinuous and the RSB scheme is 1-step with breaking parameter 
$\ol{m}=1$ at criticality. In this case there exists a (dynamical) glass transition, associated to the static one, occurring at some temperature $T_d$ greater than 
$T_{s}$. 
The static and dynamical transition temperature and overlap can be determined numerically with great accuracy using the marginality condition
and the techniques described in Ref. \cite{parisiPotts}. We briefly summarize the results here.\\
We can compute the free energy (\ref{pottsaction}) in the 1-RSB ansatz with $q_1=q$ and $q_0=0$ and expand it 
at first order around $\ol{m}=1$ as $\G_0 + (\ol{m}-1) \G_1(q)$,
\beq
\begin{split}
&\G(q) =\frac{1}{4}\beta^2(1-p)-\log(p) +(\ol{m}-1) \times \\
&\biggl (\frac{1}{4}\beta^2(p-1)q^2\,+\,\frac{1}{2}\beta^2 q(p+1)\,+\,\log(p)- I_2\biggr )
\label{eq20}
\end{split}
\eeq
\noindent where the integral $I_2$ is given by,
\beq
\begin{split}
I_2 &=\exp(-\frac{\beta^2pq}{2})\,
\int_{-\infty}^{\infty}\,\prod_{r=1}^p\,\left(\frac{dy_r}
{\sqrt{2\pi}}e^{-\frac{y_r^2}{2}} \right)\, \times \\
&e^{\beta \sqrt{qp}y_1}
\log\biggl[ \left(\sum_{r=1}^p\,exp(\beta(qp)^{\frac{1}{2}}y_r)\right)\biggr]
\label{eq21}
\end{split}
\eeq
For $m=1$ the expression (\ref{eq20}) gives the high-temperature free energy which
is independent of $q$. 
 This general expansion allows to determine the
static and the dynamic transition.\\
The static temperature is determined imposing 
that a solution $q_{s}$ exists, satisfying the following conditions:
\begin{eqnarray}
\label{saddle}
\biggl (\frac{\partial \G}{\partial q}\biggr )_{q=q_{s}}=\biggl (
\frac{\partial \G_1}{\partial q}\biggr )_{q=q_{s}}=0\\
(\G_1)_{q=q_{s}}=0
\label{eq23}
\end{eqnarray}
On the other hand, for the dynamical transition temperature, 
we must search for a marginal stability and the condition becomes.  
\begin{eqnarray}
\biggl (\frac{\partial \G}{\partial q}\biggr )_{q=q_d}=\biggl (
\frac{\partial \G_1}{\partial q}\biggr )_{q=q_d}=0\\
\biggr (\frac{\partial^2 \G}{\partial q^2}\biggr )_{q=q_d}=
\biggl (\frac{\partial^2 \G_1}{\partial q^2}\biggr )_{q=q_d}=0~~~~.
\label{eq24}
\end{eqnarray}
In the language of the Franz-Parisi potential \cite{Franz95a} the two 
conditions above correspond, respectively, to the appearence 
of a local minimum (horizontal flex) for the dynamical transition, 
and to the fact that this minimum reaches the same height of the 
paramagnetic one, for the static transition.\\
As we will see in the following, the $p$-dimensional integral $I_2$ in (\ref{eq20}) 
is extremely hard to evaluate numerically as soon as $p>2$. 
Therefore in Ref. \cite{parisiPotts} the authors use the identity 
\beq
\log(1+A)\,=\,\int_0^{\infty}\,\frac{dx}{x}e^{-x}(1-e^{-Ax})~~~~~.
\label{eq25}
\eeq
\noindent and taking
\beq
A\,=\,\sum_{r=1}^p\,\exp(\beta(qp)^{\frac{1}{2}}y_r)-1
\label{eq26}
\eeq
\noindent they obtain the result
\beq
I_2\,=\,\int_0^{\infty}\,\frac{dx}{x}e^{-x}\,\lbrace
1-e^{x}w(xe^{\beta^2qp})
w^{p-1}(x)\rbrace~~~~~~~~~.
\label{eq27}
\eeq
\noindent with
\beq
\label{wdix}
w(x)=\int_{-\infty}^{\infty} \, \frac{dy}{\sqrt{2\pi}} \, \exp \left(-\frac{1}{2} y^2 -x\exp \left(\b (pq)^{\frac{1}{2}} y\right)  \right)
\eeq
\noindent which is much easier to evaluate numerically \footnote{Note that formula (\ref{eq27}) differs from the one in ref. \cite{parisiPotts} in which 
there was a typing mistake (a $1/2$ in the argument of the exponential).}. \\

\subsubsection*{Infinite number of colours}
It has been pointed out \cite{Gross85} that the Potts 
model becomes a Random Energy Model (REM) in the limit 
$p \rightarrow \infty$, with a critical temperature that diverges like 
\beq
T_s = \frac{1}{2} \sqrt{\frac{p}{\log (p)}}
\eeq
In the following, we show that the limit model is not exactly a REM.
The first of Eq.s (\ref{saddle}), which is satisfied both at the dynamical and statical transition, can be written in the following way
\beq
\label{resaddle}
\begin{split}
q =\frac{1}{p-1} \left(p\,L^{(p)}(\b,q)-1 \right)
\end{split}
\eeq
with
\beq
\nonumber
\begin{split}
L^{(p)}(\b,q)&= \left( \int \GG_p(\ul{z}) \sum_{r=1}^p \exp \left( \b (pq)^{1/2} z_r\right)  \right)^{-1} \times \\
&\int \GG_p(\ul{z}) \left(\frac{ \sum_{r=1}^p \exp \left( 2 \b (pq)^{1/2} z_r\right) }{\sum_{r=1}^p \exp \left( \b (pq)^{1/2} z_r\right)  }\right) 
\end{split}
\eeq
and
\beq
\label{measure}
\GG_p(\ul{z})=\left( \prod_{r=1}^p \frac{dz_r}{\sqrt{2\pi}} \exp \left( - \frac{1}{2}z_r^2\right)\right)
\eeq
We have been able to show that, as a function of the rescaled temperature 
\beq
\xi=\b^2 \frac{p}{\log (p)}
\eeq
the right hand side of equation (\ref{resaddle}) tends to a Heaviside function in the infinite $p$ limit (see Appendix),
\beq
L^{(\infty)}(\xi ,q)=\theta \left(q-\frac{2}{\xi} \right)
\eeq
with breaking point $q=2/\xi$. \\
\begin{figure}
\includegraphics[width=.99\columnwidth]{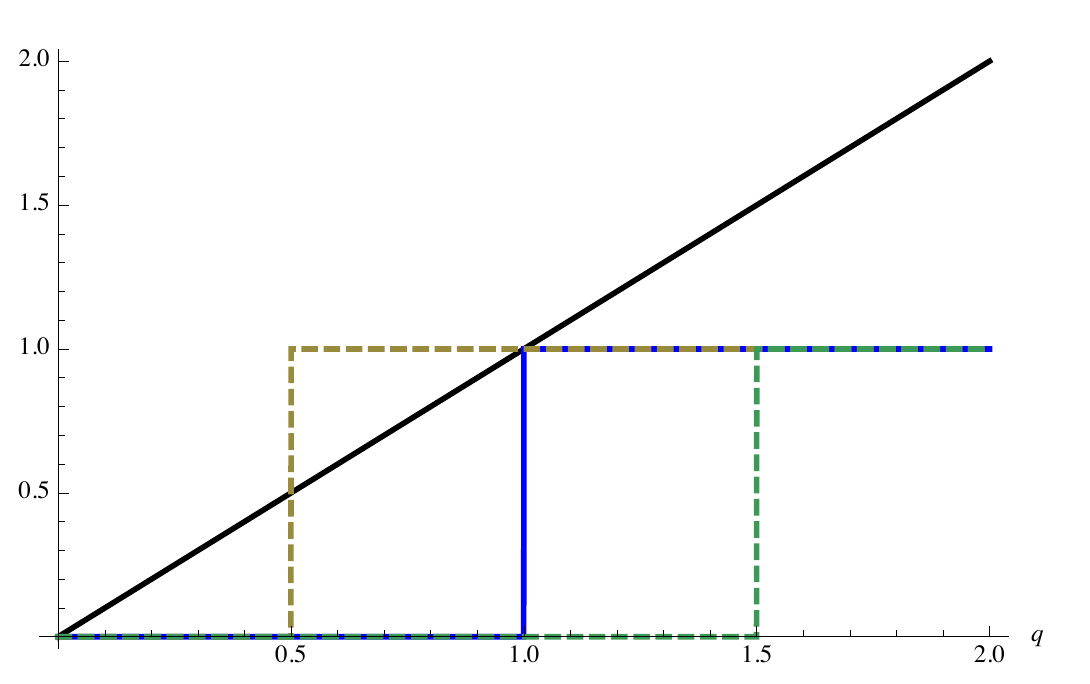}
\caption{(Color online) Black solid line: $y=q$. Yellow (left) and green (right) dashed lines: $y=L^{(\infty)}(\xi,q)$ for $\xi=4$ and $\xi=4/3$. Blue (grey) solid line: $y=L^{(\infty)}(\xi,q)$ for $\xi=2$. \\
The dynamical transition is located in $\xi=2$, for which the line $y=q$ is tangent to the curve $y=L^{(\infty)}(\xi,q)$.}
\label{fig:step}
\end{figure}
As can be easily seen from Fig. \ref{fig:step} both $q_d$ and $q_s$ go to $1$ in the limit $p\rightarrow \infty$ and 
the dynamical transition is located at the rescaled temperature such that the 
breaking point of $L^{(\infty)}$ is $1$.
\beq
\xi_d = 2\,\,\, \Longrightarrow \,\,\, T_d=\sqrt{\frac{p}{2\log (p)}} 
\eeq
While in a random energy model (REM) the ratio between $T_d$ and $T_s$ is formally infinite \cite{Derrida81}, 
in the large -$p$ Potts model this ratio tends to a finite value, namely
\beq
\frac{T_d}{T_s}\xrightarrow[p\rightarrow \infty]{} \sqrt{\frac{\xi_s}{\xi_d}}=\sqrt{2} \approx 1.414\cdots
\eeq
therefore, the limit model is still a ``glassy'' model with a dynamic and a static transition. \\
This is at variance with the Ising $p$-spin model in the $p\rightarrow \infty$ limit 
that goes to a REM \cite{Derrida81}.

\section{The Potts Model: MCT exponents}
\label{sec:pottsexp}
The determination of the mode coupling exponents follows essentially the steps 
described in section \ref{sec:how}.
Expanding the effective action (\ref{pottsaction}) to third order 
around the replica symmetric saddle point we obtain the eight coefficients
\beq
\begin{split}
& W_1=R_1 -3(p-1)qM_2 + 2(p-1)^3 q^3 \\
& W_2=R_2 -3(p-1)qM_1 + 2(p-1)^3 q^3 \\
& W_3=R_3 -(p-1)qM_1 - 2(p-1)qM_2 + 2(p-1)^3 q^3 \\
& W_4=R_4 -(p-1)qM_1 - 2(p-1)qM_3 + 2(p-1)^3 q^3 \\
& W_5=R_5 -2(p-1)qM_2 - (p-1)qM_3 + 2(p-1)^3 q^3 \\
& W_6=R_6 -3(p-1)qM_2 +2(p-1)^3 q^3 \\
& W_7=R_7 -(p-1)qM_2 - 2(p-1)qM_3 + 2(p-1)^3 q^3 \\
& W_8=R_8 -3(p-1)qM_3 + 2(p-1)^3 q^3 \\
\end{split}
\eeq
\noindent where the replica symmetric overlap is determined through the saddle point equation
\beq
q=\langle \langle \eta(\s^{\a},\s^{\b}) \rangle \rangle
\eeq
\noindent the ``mass matrix'' can assume three different values
\beq
\begin{split}
& M_1=\langle \langle \eta(\s^{\a} , \s^{\b}) \eta(\s^{\a} , \s^{\b}) \rangle \rangle \\
& M_2=\langle \langle \eta(\s^{\a} , \s^{\b}) \eta(\s^{\a} , \s^{\g}) \rangle \rangle \\
& M_3=\langle \langle \eta(\s^{\a} , \s^{\b}) \eta(\s^{\g} , \s^{\d}) \rangle \rangle \\
\end{split}
\eeq
\noindent and the six-replica cumulants are given by
\beq
\begin{split}
& R_1 = \langle \langle  \eta(\s^{\a} , \s^{\b}) \eta(\s^{\b} , \s^{\g}) \eta(\s^{\g} , \s^{\a})\rangle \rangle  \\
& R_2 = \langle \langle \eta(\s^{\a} , \s^{\b})\eta(\s^{\a} , \s^{\b})\eta(\s^{\a} , \s^{\b})\rangle \rangle\\
& R_3 = \langle \langle \eta(\s^{\a} , \s^{\b})\eta(\s^{\a} , \s^{\b})\eta(\s^{\a} , \s^{\g})\rangle \rangle \\
& R_4 = \langle \langle \eta(\s^{\a} , \s^{\b})\eta(\s^{\a} , \s^{\b})\eta(\s^{\g} , \s^{\d})\rangle \rangle\\
& R_5 = \langle \langle \eta(\s^{\a} , \s^{\b})\eta(\s^{\b} , \s^{\g})\eta(\s^{\g} , \s^{\d})\rangle \rangle \\
& R_6 = \langle \langle \eta(\s^{\a} , \s^{\b})\eta(\s^{\a} , \s^{\g})\eta(\s^{\a} , \s^{\d})\rangle \rangle \\
& R_7 = \langle \langle \eta(\s^{\a} , \s^{\b})\eta(\s^{\a} , \s^{\g})\eta(\s^{\d} , \s^{\m})\rangle \rangle \\
& R_8 = \langle \langle \eta(\s^{\a} , \s^{\b})\eta(\s^{\g} , \s^{\d})\eta(\s^{\m} , \s^{\n})\rangle \rangle \\
&\\
&\\
&\\
\end{split}
\eeq
Given the relationship (\ref{ws}) between $w_1$, $w_2$ and the $W$ coefficients one obtains: 
\beq
\begin{split}
w_1&=R_1 - 3 R_5 + 3 R_7 - R_8 \\
w_2 &= \frac{1}{2} [R_2 - 6 R_3 + 3 R_4 + 6 R_5 \\
 &+ 4 (R_6 - 3 R_7 + R_8)]
\end{split}
\eeq
where only the disconnected cumulants are left.\\
If the thermodynamic phase transition is continuous, then it coincides with the dynamical one (as in the SK model). In this 
case dynamical quantities at infinite time relax to their static value \cite{Sompolinsky82} 
and the averages above can be computed in a replica symmetric ansatz taking finally the limit $n \rightarrow 0$. 
If, instead, the transition is discontinuous then the coefficients have to be computed at the dynamical transition, where 
quantities at infinite time do {\it not} relax to their equilibrium (thermodynamic) value but remain stuck at their value inside 
the most excited metastable states.
The averages should then be computed {\it inside} a single state; this corresponds to taking a 1-RSB ansatz with 
breaking parameter $m \rightarrow 1$ or, if the mutual overlap between different states is $0$ as in our case, a RS ansatz 
with the number of replicas $n \rightarrow 1$ \cite{Crisanti08}.
Finally, we can assume replica symmetry and leave $n$ unspecified, obtaining 
the expression for the two coefficients.
\beq
w_1= p^3 (L_3- 3 L_4 + 3 L_{23} -L_{222})
\label{coeffpotts}
\eeq
and
\beq
\nonumber
\begin{split}
& w_2  =\frac{p^2}{2} \left( 1 - q\right) +  \\
& + \frac{1}{2} p^3 \left( q - 6 L_3 + 10 L_4 + 3 L_{22} -12 L_{23} + 4 L_{222}\right)
\end{split}
\eeq
\noindent where, exploiting the fact that $\left(\eta(\s^{a} , \s^{b})\right)^2 = (p-1) + (p-2) \eta(\s^{a} , \s^{b})$ the saddle point
equation becomes

\begin{widetext}

\beq
\begin{split}
q=\frac{1}{p-1}\left[ p \frac{\int \GG_p(\ul{z}) \left( \sum_{r=1}^p \exp \left( \b (pq)^{1/2} z_r\right) \right)^{n-2} \left( \sum_{r=1}^p \exp \left( 2 \b (pq)^{1/2} z_r\right)\right)}{\int \GG_p(\ul{z}) \left( \sum_{r=1}^p \exp \left( \b (pq)^{1/2} z_r\right) \right)^{n} } - 1 \right]
\end{split}
\eeq
\noindent and we have defined the class of integrals
\beq \begin{split}
 &L_{klh}  = \frac{1}{ \int \GG_p(\ul{z}) \left( \sum_{r=1}^p \exp \left( \b (pq)^{1/2} z_r\right) \right)^{n} } 
\times \int \GG_p(\ul{z}) \left( \sum_{r=1}^p \exp \left( \b (pq)^{1/2} z_r\right) \right)^{n-k-l-h} \times \\
& \times\left( \sum_{r=1}^p \exp \left( k \b (pq)^{1/2} z_r\right)\right) \times \left( \sum_{r=1}^p \exp \left( l \b (pq)^{1/2} z_r\right)\right)  
\times \left( \sum_{r=1}^p \exp \left( h \b (pq)^{1/2} z_r\right)\right)     \\
& \\
&L_{kl}=\frac{1}{p} L_{kl0}\\
& \\
&L_{k} = \frac{1}{p^2} L_{k00} 
\label{horror}
\end{split}
\eeq

\end{widetext}

\noindent with $\GG_p(\ul{z})$ given in Eq. (\ref{measure}).\\
As already pointed out, the above result holds both for continuous and discontinuous transitions with the 
only difference that in the former case $q$ and the $L$ integrals are computed at $n=0$ while in the latter 
we consider $n=1$.

\subsection{The continuous transition}
If $p<4$ the phase transition is second-order \cite{Gross85} with $q(x)$ continuous for $p=2$ and
step-like for $p=3$. 
In the case of continuous transitions we have to consider $n=0$ and $q=0$ and the 
result (already found in Ref. \cite{Gross85}) is very simple, namely:
\beq
\label{cont}
\frac{w_2}{w_1}=\frac{p-2}{2}
\eeq
which yields $\nu_2=0.5$ and $\nu_3\simeq 0.395$.\\
As in the case of the fully-connected model, it can be proven \cite{Goldschmidt88} that, on the Bethe lattice, for $p\leq4$ the 
phase transition is second-order. The difference on the Bethe lattice is that, for $p=3$ and low enough connectivity, the order parameter $q(x)$ is a 
continuous (Parisi type) function while for high connectivity it becomes a step-like function (as in the fully-connected case). This does not affect the result which is 
again (for $p\leq 4$) given by Eq. (\ref{cont}).

\subsection{The discocontinuous transition}
We are interested here in the case $p>4$, when the system undergoes a dynamical transition, therefore we must take the limit 
$n\to 1$ in order to compute the exponents correctly.\\
The computation of the overlap $q$ and of the third order coefficients $w_1$ and $w_2$ involve $p$-dimensional integrals (\ref{horror}), which become 
very difficult to evaluate numerically as soon as $p$ is greater than $2$. In order to overcome this issue, using the identity\\
\beq
\label{identity}
\frac{1}{A^{k}}=\frac{1}{(k-1)!} \int_0^{\infty} \, x^{k-1} e^{-Ax}
\eeq
the integrals (\ref{horror}) can be rewritten in following form, which is more suitable for numerical evaluation

\begin{widetext}

\beq
\label{megaint}
\begin{split}
& L_{klh} =\frac{e^{-\frac{1}{2}\b^2 p q}}{p\, (k+l+h-2)!} [p  \, e^{\frac{1}{2}(k+l+h)^2\b^2 pq} \int_0^{\infty} \, dx \, x^{k+l+h-2} \, w^{p-1}(x) \, w\left( x\,  e^{(k+l+h)\b^2 pq} \right) + \\
& + p(p-1)  \, e^{\frac{1}{2}(k+l)^2\b^2 pq} e^{\frac{1}{2}h^2\b^2 pq}\int_0^{\infty} \, dx \, x^{k+l+h-2} \, w^{p-2}(x) \, w\left( x\,  e^{(k+l)\b^2 pq} \right) w\left( x\,  e^{h\b^2 pq}\right) + \\
& + p(p-1)  \, e^{\frac{1}{2}(k+h)^2\b^2 pq} e^{\frac{1}{2}l^2\b^2 pq}\int_0^{\infty} \, dx \, x^{k+l+h-2} \, w^{p-2}(x) \, w\left( x\,  e^{(k+h)\b^2 pq} \right) w\left( x\,  e^{l\b^2 pq}\right) + \\ 
& + p(p-1)  \, e^{\frac{1}{2}(l+h)^2\b^2 pq} e^{\frac{1}{2}k^2\b^2 pq}\int_0^{\infty} \, dx \, x^{k+l+h-2} \, w^{p-2}(x) \, w\left( x\,  e^{(l+h)\b^2 pq} \right) w\left( x\,  e^{k\b^2 pq}\right) + \\
& + p(p-1)(p-2)  \, e^{\frac{1}{2}(k^2+l^2+h^2)\b^2 pq} \int_0^{\infty} \, dx \, x^{k+l+h-2} \, w^{p-3}(x) \, w\left( x\,  e^{k\b^2 pq} \right) w\left( x\,  e^{l\b ^2 pq}\right) w\left( x\,  e^{h\b^2 pq} \right)\\
\end{split}
\eeq

\end{widetext}

\noindent with $w(x)$ given in equation (\ref{wdix}).\\
Through identity (\ref{identity}) we have been able to reduce $p$-dimensional to ``sort of'' $2$-dimensional integrals. \\
They are not technically $2$-dimensional integrals because in formula (\ref{megaint}) for each value of the integration variable 
$x$ we have to perform $2$, $3$ or $4$ integrations to obtain the function $w$ in different points (instead of just one integration 
which would be needed in a regular $2$-d integral). \\
Some care is needed in the computation of $w(x)$, especially for small $x$'s since the integrand 
has an extremely steep growth near zero and the integration step must be taken very small. \\
In order to go to very large values of $p$ it should be better to recast the integrals (\ref{megaint}) in their asymptotic form 
using Eq.s (\ref{seconda}) and (\ref{wtothep}) given in the Appendix.
The results for different values of the number of colours $p$ are reported in the table below and in Fig. \ref{fig:esponente}.

\begin{center}
\begin{table}
  \begin{tabular}{| c | c | c | c | c |}
    \hline
    $p$  & $T_{d}$ & $q_d$ & $\l$ & $a$  \\ \hline
    5  & 1.0101 & 0.09507 & 0.8764 & 0.2290  \\ \hline
    7  & 1.0577 & 0.2206 & 0.8236 & 0.2651  \\ \hline
    10 & 1.1420 & 0.3238 & 0.8052 &  0.2759  \\ \hline
    12 & 1.1970 & 0.3665 & 0.8002 &  0.2787  \\ \hline
    15 & 1.2748 & 0.4114 & 0.7962 &  0.2810  \\ \hline
    20 & 1.3926 & 0.4598 & 0.7930 & 0.2827  \\ \hline
    30 & 1.5941 & 0.5142 & 0.7904 & 0.2841  \\ \hline
    40 & 1.7648 & 0.5455 & 0.7895 & 0.2846  \\ \hline
    100 & 2.4964 & 0.6187 & 0.7892 & 0.2848  \\ \hline
  \end{tabular}
  \caption{The dynamical transition temperature, the dynamical overlap, the exponent parameter and the $a$ exponent 
  for different values of $p$ ranging from 5 to 100}
  \end{table}
\end{center}

\section{Comparison with numerical simulations}
\label{sec:simul}
For $p=10$ there are Monte Carlo simulations performed in Ref. \cite{Brangian01, Brangian02bis} to investigate the finite-size effects 
on the glass transition. 
They find that the thermodynamic static quantities such as the energy, 
the entropy, the susceptibility and the overlap distribution display 
very strong finite size effects.
It is found also that the system remains always ergodic and the {\it plateau} in the equilibrium spin-spin correlation function $C(t)$  is almost 
invisible even at temperarures close to the dynamical transition $T_G$ and 
big system sizes.
Since for $N \rightarrow \infty$ the physics of the system should be descibed by the exact mode-coupling equations, 
they expect a divergence of the relaxation time $\t(t)$ with a power law behaviour at the dynamical transition. 
\beq
\t \propto \left(\frac{T}{T_d}-1 \right)^{-\g}
\eeq
with an exponent $\g$ which, in mode coupling theory is 
related to the exponents $a$ and $b$ through the exact relation 
\beq
\label{relgamma}
\g=\frac{1}{2 a} + \frac{1}{2 b}
\eeq
They plot $\t^{-\frac{1}{\g}}$ for a set of 
reasonable trial values of $\g$ and find that the data are
linearized in the region $1.1\leq T \leq 1.4$ for $\g=2.0 \pm 0.5$. 
This value of $\gamma$ gives, through the relation (\ref{relgamma}), an indirect estimate 
for $a(\gamma)\approx 0.36$.
Two different kinds of finite-size scaling are considered in order to perform extrapolations of $C(t,N)$ at $N\rightarrow \infty$.
They find that only one of the two 
gives a $C(t)$ which is well compatible with a power 
law behaviour of the type (\ref{eqa}). 
In this way they can make a rough direct estimate of the 
exponent, obtaining $a=0.33 \pm 0.04$ which,
despite the difficulties (identification of the plateau, extrapolation etc...) is close (within $2\s$) to our exact computation.

\section{Conclusions and remarks}
\label{sec:conclusion}

\begin{figure}
\includegraphics[width=.99\columnwidth]{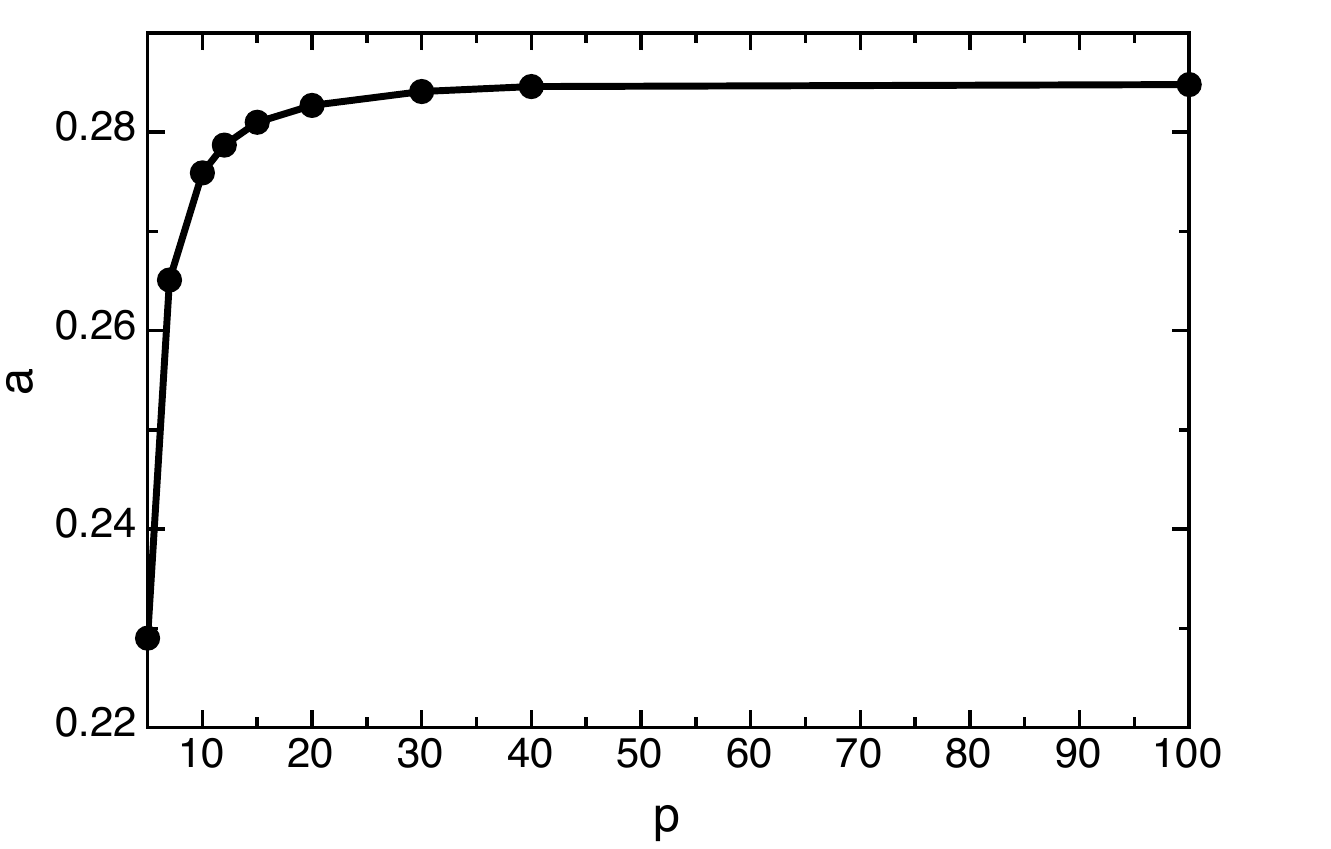}
\caption{Black joined circles: the exponent $a$ computed for different 
values of the number of colours ranging from $p=5$ to $p=100$ in the discontinuous regime.}
\label{fig:esponente}
\end{figure}

In the first part of this paper we have presented a review of some known 
results about the disordered Potts model. In the second part, exploiting a technique that 
has been recently developed \cite{Calta11}, we have 
computed the dynamical exponents of the autocorrelation decay both in the case of continuous 
and discontinuous transition, in a completely static framework.\\
In Fig. \ref{fig:esponente} we show the 
plot of the exponent $a$ as a function of the 
parameter $p$ from $p=5$ up to $p=100$.

Since our computation is non-perturbative, the exponents 
can be determined with arbitrary precision and they 
can be taken as a reference in numerical simulations. 
Knowing a priori the exponents in the thermodynamic limit, one has an additional tool for
studying, for example, the finite size effects and the deviations from MCT in a numerical simulation of a finite (fully-connected) system.

\subsection*{Acknowledgements}
The authors would like to thank Luca Leuzzi and Ulisse Ferrari for 
many useful discussions and for a critical reading of the manuscript.

\appendix

\section{Infinite $p$ limit}
\begin{figure}
\includegraphics[width=.99\columnwidth]{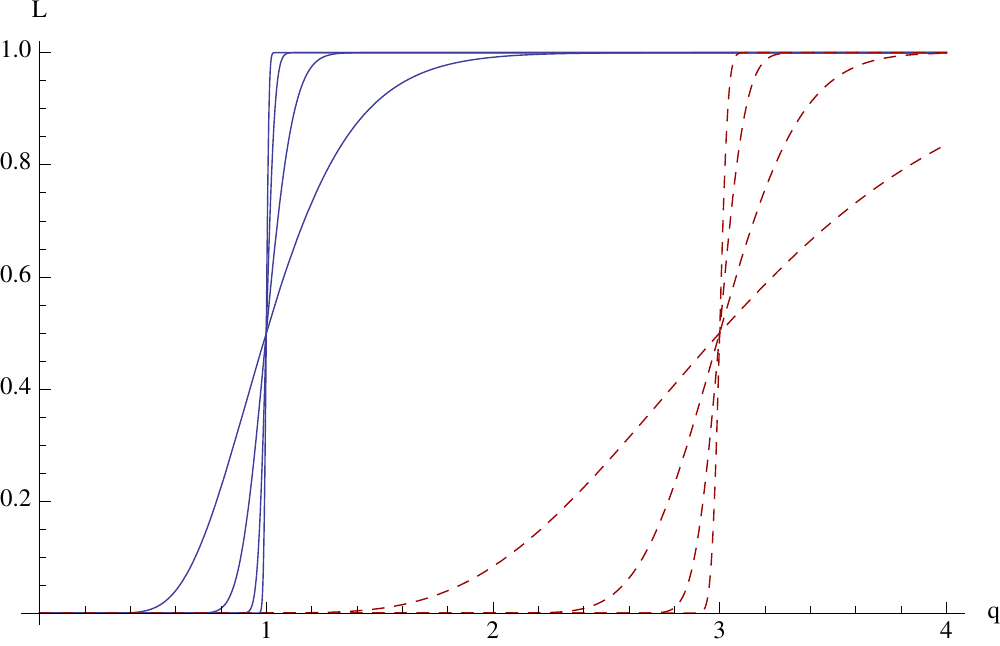}
\caption{(Color online) The right hand side of equation (\ref{approx}) for $p=10^{10}, 10^{10^2}, 10^{10^3}, 10^{10^4}$.\\
Solid blue lines: $\xi=2$. Dashed orange lines: $\xi=2/3$.}
\label{fig:steps}
\end{figure}

In this Appendix we compute the dynamical critical temperature $T_d$ and overlap $q_d$ in the limit $p\rightarrow \infty$ and the first correction.\\
First of all, note that the saddle point equation (\ref{resaddle}) in the infinite $p$ limit becomes 
\beq
q=L^{(\infty)}(\b,q)
\eeq
Using identity (\ref{identity}) the $p$-dimensional integral $L$ of Eq. (\ref{resaddle}) can be rewritten in the following way
\beq
\label{ellep}
L^{(p)}(\b,q)=\int_0^{\infty} dx \left[w\left( x e^{-\frac{3}{2}\b^2 p q}\right)\right]^{p-1} w\left( x e^{\frac{1}{2}\b^2 p q}\right)
\eeq
Setting $\a \equiv \b p^{1/2}q^{1/2}$ and making the change of variables 
\beq
x=e^{\frac{1}{2} \a ^2 + \a \ox}
\eeq
the integral becomes 
\beq
\label{prima}
\int_{-\infty}^{\infty} d\ox \, \a \, e^{\frac{1}{2} \a ^2 + \a \ox} w\left( e^{\a^2 + \a \ox}\right)\left[w\left( e^{-\a^2 +\a \ox}\right)\right]^{p-1} 
\eeq
Through standard manipulations it can be shown that
\beq
\label{seconda}
\a \, e^{\frac{1}{2} \a ^2 + \a \ox} w\left( e^{\a^2 + \a \ox}\right) = \frac{1}{\sqrt{2 \p}} e^{-\frac{1}{2} \ox ^2} \,F(\ox)
\eeq
where $F(\ox)$ tends to unity in the $\a \rightarrow \infty$ limit
\beq
F(\ox)=\int_{-\infty}^{\infty} \, dy \, \left[ e^{y-e^y}\right] e^{-\frac{y^2}{2\a^2}+\frac{y\ox}{\a}}
\eeq
Moreover we have clearly
\beq
w\left( e^{-\a^2 +\a \ox}\right)\rightarrow 1
\eeq
The behaviour of $w^{p-1}$ and, consequently, of the integral $L$ is now determined by the leading order of the first correction  $\D \equiv w-1$, in fact 
\beq
\label{wtothep}
\left[w\left( e^{-\a^2 +\a \ox}\right)\right]^{p-1} \simeq \exp \left[ p\,\log \left(1+\D \right)\right]
\eeq
We have to compute the leading behaviour of
\beq
\D \equiv \int_{-\infty}^{\infty}\,\frac{dy}{\sqrt{2\pi}}\, e^{-\frac{1}{2} y^2} \left[ \exp \left(- e^{-\a^2 + \a (y+\ox)} \right) -1\right]
\eeq
making the change of variables
\beq
y=z+\a -\ox
\eeq
we obtain the form
\beq
\D = e^{-\frac{1}{2} \a ^2 +\a \ox}  \int_{-\infty}^{\infty}\,\frac{dz}{\sqrt{2\pi}}\, e^{-\frac{1}{2} (z-\ox)^2} e^{-\a z}\left[ \exp \left( -e^{\a z} \right) -1\right]
\eeq
which in the limit of large $\a$ goes to the following form
\beq
\begin{split}
\D &\simeq - e^{-\frac{1}{2} \a ^2 +\a \ox}  \int_{-\infty}^{\infty}\,\frac{dz}{\sqrt{2\pi}}\, e^{-\frac{1}{2} (z-\ox)^2} \theta(-z)\\
& = -\frac{1}{2} e^{-\frac{1}{2} \a ^2 +\a \ox} \, \text{Erfc}\left( \frac{\ox}{\sqrt{2}} \right) 
\end{split}
\eeq
Given this correction, Eq. (\ref{wtothep}) becomes
\beq
\label{wtothep2}
w^{p-1} \simeq \exp \left[-\frac{p}{2} e^{-\frac{1}{2} \a ^2 +\a \ox} \, \text{Erfc}\left( \frac{\ox}{\sqrt{2}} \right) \right]
\eeq
Using a rescaled inverse temperature $\b^2 = \xi \log(p)/p$ we have
\beq
\label{subs}
\a ^2 = \xi q \log (p)
\eeq
and substituting into Eq. (\ref{wtothep2}) one obtains the following expression
\beq
\label{wtothep3}
w^{p-1} \simeq \exp \left[-\frac{1}{2} p^{1-\frac{\xi q}{2}} e^{ \ol{x} \sqrt{\xi q \log (p)}} \, \text{Erfc}\left( \frac{\ox}{\sqrt{2}} \right) \right]
\eeq
Independently of the value of $\ol{x}$, the quantity (\ref{wtothep3}) clearly goes to $1$ if $\xi q > 2$ while 
it goes to $0$ if $\xi q < 2$. Therefore, the integrand in Eq. (\ref{ellep}) converges {\it uniformly} to a normalized gaussian if $q > 2/\xi$ or 
to $0$ if $q < 2/\xi$ and, since the convergenge is uniform, the limit can be taken {\it before} the integration, yielding (see Fig. \ref{fig:steps})
\beq
\begin{split}
L^{(\infty)}(\xi,q) &\equiv \lim_{p\rightarrow \infty} L^{(p)}\left(\sqrt{\xi \frac{\log(p)}{p}} ,q\right)\\
&=\theta\left( q-\frac{2}{\xi} \right)
\end{split}
\eeq
The dynamical transition will be located at the temperature for which the two curves $y=q$ and $y=L^{(\infty)}(\xi,q)$ are tangent, that 
is, when the braking point of the step function is $1$. For this reason we have 
\beq
\begin{split}
&q_d^{(\infty)}=1\\
&\xi_d^{(\infty)}=2
\end{split}
\eeq
We have obtained the desired result in the infinite $p$ limit and now we 
want to compute the leading correction both to the critical temperature and 
to the critical overlap starting again from equation (\ref{wtothep2}).\\
In order to obtain a finite result for finite $\ox$ we must have: 
\beq
\label{condition}
\log (p) - \frac{\a ^2}{2} + t \a = 0
\eeq
for some value of $t$ wich now becomes our variable. \\
Under this condition $w^{p-1}$ behaves like $\theta(t-\ox)$ and the 
equation for the overlap now reads
\beq
\label{qlim}
q=\frac{1}{2} \left(1+\text{Erf}\left( \frac{t}{\sqrt{2}} \right)\right)
\eeq
and Eq. (\ref{condition}) with the substitution (\ref{subs}) becomes
\beq
\label{xilim}
\xi q - \frac{2t}{\sqrt{\log(p)}} (\xi q)^{\frac{1}{2}} -2 =0
\eeq
Substituting Eq. (\ref{xilim}) into Eq. (\ref{qlim}) we get
\beq
\label{approx}
q=\frac{1}{2} \left(1+\text{Erf}\left( \frac{1}{2\sqrt{2}} (\xi q -2)\sqrt{\frac{\log(p)}{\xi q}} \right)\right)
\eeq
Eq. (\ref{approx}) can be used to obtain approximate solutions in the large $p$ limit.\\
Considering that $1<< t << \log(p) $, from Eq.s (\ref{qlim}) and (\ref{xilim}) , we have at leading order
\beq
\label{eqlimite}
\begin{split}
&q\simeq 1-\frac{e^{-t^2/2}}{t\sqrt{2\p}}\\
&\xi q  \simeq 2 + 2\sqrt{2} \frac{t}{\sqrt{\log(p)}} + ..
\end{split}
\eeq
we can then define
\beq
\begin{split}
&\e = 1-q \\
&\mu = \xi -2
\end{split}
\eeq
satisfying the two coupled equations
\beq
\label{coupled}
\begin{split}
&\e  =\frac{e^{-t^2/2}}{t\sqrt{2\p}}\\
&\mu - 2\e= 2 \sqrt{2}\frac{t}{\sqrt{\log(p)}}
\end{split}
\eeq
We can obtain $t^2$ from the second of Eq.s (\ref{coupled}) and plug it 
into the logarithm of the first one getting, at leading order 
\beq
\label{epsilonsaddle}
\log(\e)=-\frac{1}{16} \log(p) \left( \mu - 2 \e\right)^2
\eeq
At criticality the derivative of Eq. (\ref{epsilonsaddle}) must hold as well, 
giving the second contraint necessary to determine both $\e$ and $\mu$
\beq
\label{epsilonsaddle2}
\frac{1}{\e}=\frac{1}{4} \left( \mu -  2\e \right) \log(p)
\eeq
Substituting this last equation into (\ref{epsilonsaddle}) one obtains
\beq
\e ^2\log(\e)=-\frac{1}{\log(p)}
\eeq
which at leading order gives
\beq
\e=\left[ \frac{2}{\log(p) \log\left(\log(p)\right)} \right]^{\frac{1}{2}}
\eeq
Substituting into (\ref{epsilonsaddle2}) and taking the leading order we get the other correction
\beq
\begin{split}
\m=2\sqrt{2}\left[ \frac{ \log\left(\log(p)\right)}{\log(p)} \right]^{\frac{1}{2}}\\
\end{split}
\eeq
Given $\e$ and $\mu$ we can write the dynamical overlap and critical temperature 
with the leading correction for large $p$:
\beq
\begin{split}
&q_d=1-\left[ \frac{2}{\log(p) \log\left(\log(p)\right)} \right]^{\frac{1}{2}}\\
&T^2_d=\frac{p}{2\log(p)} \left[ 1- \sqrt{2}\left[ \frac{ \log\left(\log(p)\right)}{\log(p)} \right]^{\frac{1}{2}}  \right]
\end{split}
\eeq
\\
\\
\\
\\
\\
\\
\\
\\

\bibliography{francescobib.bib}

\end{document}